\documentclass[twocolumn,twocolumn]{IEEEtran}
\usepackage[T1]{fontenc}
\usepackage{color}
\usepackage{float}
\usepackage{mathrsfs}
\usepackage{amsmath}
\usepackage{amssymb}
\usepackage{graphicx}

\graphicspath{{figures/}}
\usepackage[unicode=true,
 bookmarks=false,bookmarksnumbered=true,bookmarksopen=true,bookmarksopenlevel=1,
 breaklinks=false,pdfborder={0 0 0},pdfborderstyle={},backref=false,colorlinks=false]
 {hyperref}
\hypersetup{pdftitle={Your Title},
 pdfauthor={Your Name},
 pdfpagelayout=OneColumn, pdfnewwindow=true, pdfstartview=XYZ, plainpages=false}

\makeatletter

\floatstyle{ruled}
\newfloat{algorithm}{tbp}{loa}
\providecommand{\algorithmname}{Algorithm}
\floatname{algorithm}{\protect\algorithmname}

\usepackage[caption=false,font=footnotesize]{subfig}
\usepackage{algorithm}
\usepackage{algorithmic}

\usepackage{multirow} 
\usepackage{amsmath} 
\usepackage{xcolor}

\allowdisplaybreaks[4]

\ifCLASSOPTIONcompsoc
\usepackage[caption=false,font=normalsize,labelfont=sf,textfont=sf]{subfig}
\else
\usepackage[caption=false,font=footnotesize]{subfig}
\fi
\usepackage{booktabs}
\usepackage{cite}
\usepackage{bm}
\usepackage{algorithmic}
\usepackage{algorithm}
\usepackage{graphicx}
\IEEEoverridecommandlockouts

\usepackage{lettrine}

\usepackage{geometry}
\@ifundefined{showcaptionsetup}{}{%
 \PassOptionsToPackage{caption=false}{subfig}}
\usepackage{subfig}
\makeatother
\usepackage{cases}
\begin{document}

\title{\textcolor{black}{Distributionally Robust Optimization for Computation Offloading in Aerial Access Networks}}
\author{
\IEEEauthorblockN{Guanwang Jiang$^{\dagger}$, Ziye Jia$^{\dagger}$$^{\ast}$, Lijun He$^{\ddag}$, 
Chao Dong$^{\dagger}$, Qihui Wu$^{\dagger}$, and Zhu Han$^{\S}$
\\
 }
\IEEEauthorblockA{
$^{\dagger}$Key Laboratory of Dynamic Cognitive System of Electromagnetic Spectrum Space, Ministry of Industry and 
Information Technology, Nanjing University of Aeronautics and  Astronautics, Nanjing, Jiangsu, 211106, China\\
$^{\ast}$National Mobile Communications Research Laboratory, Southeast University, Nanjing, Jiangsu, 211111, China\\
$^{\ddag}$School of Software, Northwestern Polytechnical University, Xi'an, Shaanxi, 710072, China \\
$^{\S}$Department of Electrical and Computer Engineering, University of Houston, Houston, TX77004, USA\\
\{jiangguanwang, jiaziye, dch, wuqihui\}@nuaa.edu.cn, lijunhe@nwpu.edu.cn, hanzhu22@gmail.com
}
}
\maketitle
\pagestyle{empty} 

\thispagestyle{empty}
\begin{abstract}
With the rapid increment of multiple users for data offloading and computation, 
 it is challenging to guarantee the quality of service (QoS) in remote areas. To deal with the challenge, 
it is promising to combine aerial access networks (AANs) with multi-access edge computing (MEC) equipments 
to provide computation services with high QoS.  
However, as for uncertain data sizes of tasks, it is intractable to optimize the offloading decisions 
and the aerial resources. 
Hence, in this paper, we consider the AAN to provide MEC services for uncertain tasks. 
 Specifically, we construct the uncertainty sets based on historical data
to characterize the possible probability distribution of the uncertain tasks. 
Then, based on the constructed uncertainty sets, 
we formulate a distributionally robust optimization problem to minimize the system delay. 
Next,
we relax the problem and reformulate it into a linear programming problem.
 Accordingly, we design a MEC-based distributionally robust latency optimization algorithm.
 Finally, simulation results reveal that the proposed algorithm achieves 
a superior balance between reducing system latency and minimizing energy consumption, 
as compared to other benchmark mechanisms in the existing literature.
\end{abstract}

\section{INTRODUCTION}
\lettrine[lines=2]{D}UE to the development of the sixth generation communication technology (6G), 
multiple demands for data offloading  grow rapidly, such as the Internet of things (IoT) requirements \cite{5:1}. 
However, terminal equipments in remote areas are very difficult to be
 solely served by 
terrestrial cellular networks \cite{13:120}.
As important components of space-air-ground networks, the aerial access networks (AANs) can provide effective services on remote areas through unmanned aerial vehicles (UAVs), high altitude platforms (HAPs),
 as well as satellite networks. AANs have the advantages of convenient deployment, high flexibility, and low latency.
Moreover, there exist a large amount of computation tasks in the 6G AANs, such as 
farmland information and crop growth dynamicsthe in the intelligent agriculture \cite{10032258}.
 Therefore, it is necessary to integrate the multi-access edge computing (MEC) technique into 6G AAN to reduce the system latency \cite{5:2}.
On account of the small size, easy deployment, and  strong maneuverability, 
UAVs equipped with computation resources are 
introduced as effective MEC candidates \cite{jia2020leo}.
However, the resource and  endurance time of UAVs are 
constrained by their physical structures, thereby leading to the decrement of quality of service (QoS) \cite{10:23}. 
 To this end, HAPs are served as stable aerial stations and provide services 
for terrestrial devices (TDs) and UAVs, due to their large coverage and powerful payload. 
Hence, it is promising to
provide services with high quality for TDs via the cooperation of UAVs and HAPs \cite{50:2}.

There exist some recent works related with MEC in AAN. For instance, 
the authors of \cite{24} minimized the energy consumption of mobile devices and UAVs,
to devise the strategies of the task offloading, communication, and computation resource 
allocation. 
The work in \cite{45} minimized the  energy consumption of users' equipment during task offloading 
in a UAV-assisted MEC system.
 \cite{10599389} proposes the cooperative cognitive dynamic system to optimize the 
management of UAV swarms to ensure real-time adaptability. 
 In \cite{10559211}, the authors design a
trajectory planning and communication design in intelligent collaborative air-ground communication algorithm
to ensure timely information transmission among all ground users. 
 The authors in \cite{5} optimized the access scheme and power allocation in the UAV-based MEC, 
considering the errors caused by various perturbations in realistic circumstances. 
The works mentioned above are limited in practical applications, 
since they assumed that the data sizes of all the tasks are known in advance.
However, in most practical cases, the data sizes of tasks are uncertain, which affect the reliability and QoS of the system.
Traditional uncertainty optimization methods such as stochastic optimization (SO) 
and robust optimization (RO) can be employed to deal with the uncertainty problems.
However, the definite probability distribution is necessary in SO,
 while the solutions of RO are too conservative \cite{beyer2007robust}.
Different from SO, the distributionally robust optimization (DRO) 
characterizes the uncertainty based on the historical data with unknown probability distribution \cite{37:14}.
Moreover, compared to RO, DRO has a higher data utilization rate, making solution better in practice.
In view of these, DRO is a good choice for the robust offloading problem in AANs.

In this paper, we focus on the computation offloading in the AAN to reduce the system latency.
 To deal with the uncertain data sizes of tasks, 
we construct the uncertainty sets based on historical information.
 Then, the DRO problem is formulated to minimize the worst-case expected system latency. 
However, this problem is an NP-hard mixed integer minimax optimization.
To alleviate the complexity, we relax binary decision variables to reformulate 
the original problem into a linear programming form.
Accordingly, the MEC-based distributionally robust latency optimization algorithm (MDRLOA) is proposed.
Finally, simulations are constructed to verify the performance.

The rest of this paper is arranged as follows. 
Section \ref{section2} presents the system model and problem formulation. 
In Section \ref{section3}, we present MDRLOA  to
solve the proposed problem. Simulation results and discussions are provided in Section \ref{section4}.
 Finally, we draw the conclusions in Section \ref{section5}.

\begin{figure}[t]
    \centering
    \includegraphics[width=0.6\linewidth]{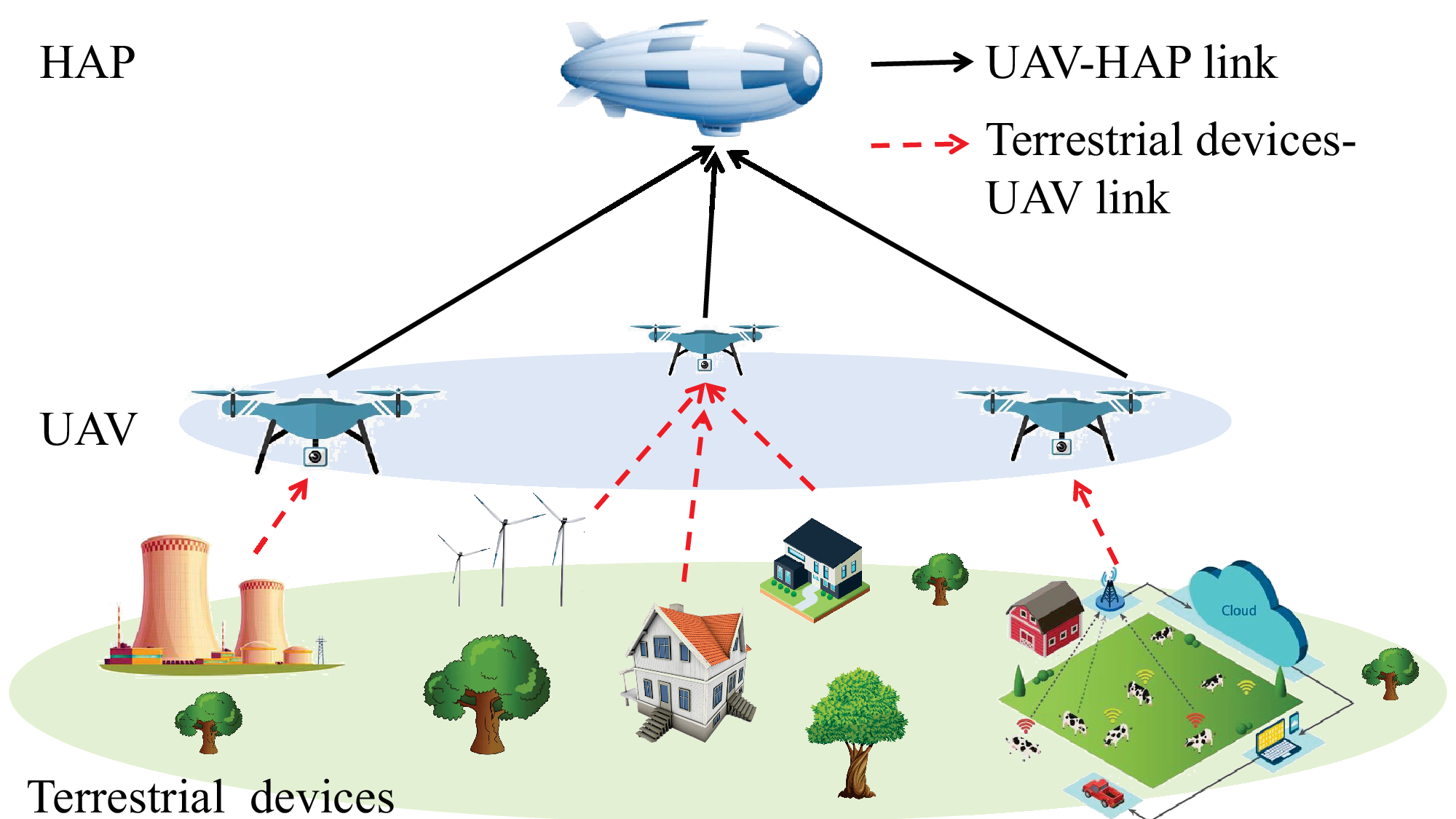}
    \caption{Computation offloading in an AAN.}
    \label{fig:Scenario}
\end{figure}

\section{SYSTEM MODEL AND PROBLEM FORMULATION}\label{section2}
As shown in Fig. \ref{fig:Scenario}, we consider an AAN in the remote area,
 composed of UAVs and a HAP in the air, and TDs on the ground. Both UAVs and HAP carry servers to 
provide computing services for TDs. In detail, the
TDs, UAVs and HAP are denoted as $i\in\mathcal{I}=\{1,2,...,I\}$, 
$j\in\mathcal{J}=\{1,2,...,J\}$, and  $H$. 
We assume the locations of the TDs, UAVs, and  HAP are fixed in the system, and  the three-dimensional Cartesian coordinate 
 is 
employed to describe the locations, denoted as $L_i^D=\left(x_i^D,y_i^D,0\right)$,  $L_j^V=\left(x_j^V,y_j^V,h_j^V\right)$ 
and  $L^H=\left(x^H,y^H,h^H\right)$, respectively. In the model, UAVs are responsible for collecting tasks from TDs, and  
then computing locally on the UAV or transmitting to HAP for further processing. 
Moreover, the accommodated devices by a UAV or HAP cannot exceed
the quota $N_u$ of UAV  or the quota $N_H$ of HAP. 
The size of the task generated by TD $i$ is denoted as $\phi_i$. 
We define the decision variables $x_{i,j}$, $y_{i,j}$, and  $z_{i,j,H}$ as 
\begin{equation}
x_{i,j}=\begin{cases}
1,&\mathrm{\textrm{if $\phi_i$ is offloaded to UAV $j$,}}\\0,&\mathrm{\textrm{otherwise,}}
\end{cases}\label{cons: x01}
\end{equation}
\begin{align}
y_{i,j}=\begin{cases}
1,&\mathrm{\textrm{if $\phi_i$ is computed by UAV $j$,}}\\0,&\mathrm{\textrm{otherwise,}}
\end{cases}\label{cons:y01}
\end{align}
and
\begin{align}
z_{i,j,H}=\begin{cases}
1,&\mathrm{\textrm{if $\phi_i$ is relayed from UAV $j$ to HAP $H$,}}\\0,&\mathrm{\textrm{otherwise.}}
\end{cases}\label{cons:z01}
\end{align}

\subsection{ Communication Model}
\subsubsection{TD to UAV Transmission}
The TD-to-UAV link is considered as line-of-sight \cite{10}, 
and the channel gain from TD $i$ to UAV $j$ is calculated as
\begin{align}
g_{i,j}=g^0_{i,j} d_{i,j}^{-2},
\end{align}
where $g^0_{i,j}$ represents the channel gain at the reference distance, and  $d_{i,j}$ is the distance between TD $i$ and UAV $j$,
  which can be calculated as
\begin{equation}
d_{i,j}=\sqrt{\left(x^D_i-x^V_j\right)^2+\left(y^D_i-y^V_j\right)^2+\left(h^V_j\right)^2}.
\end{equation}
Hence, according to Shannon theory, the data transmission rate from TD $i$ to UAV $j$ is
\begin{equation}
R_{i,j}=B_{i,j}\log_2\left(1+\frac{p^{tr}_i g_{i,j}}{\sigma^2}\right),
\end{equation}
where $B_{i,j}$ is the channel bandwidth between TD $i$ and UAV $j$, $p^{tr}_i$ is the maximum transmission power of TD $i$,
 and 
$\sigma^2$ is the noise power. Hence, the time cost to transmit data $\phi_i$ from TD $i$ to UAV $j$ is
\begin{equation}
T^{tr}_{i,j}=\frac{\phi_ix_{i,j}}{R_{i,j}}.
\end{equation}

\subsubsection{UAV to HAP Transmission}
Similarly, the UAV-to-HAP link is line-of-sight \cite{50}. 
 We denote $g^0_{j,H}$ as the channel gain at the reference distance
\begin{align}
g_{j,H}=g^0_{j,H} d_{j,H}^{-2},
\end{align}
where $d_{j,H}$ is the distance between UAV $j$ and HAP $H$,
 calculated as
\begin{equation}
d_{j,H}=\sqrt{\left(x^V_j-x^H\right)^2+\left(y^V_j-y^H\right)^2+\left(h^V_j-h^H\right)^2}.
\end{equation}
Thus, the data transmission rate from UAV $j$ to HAP $H$ is
\begin{equation}
R_{j,H}=B_{j,H}\log_2\left(1+\frac{p^{tr}_j g_{j,H}}{\sigma^2}\right),
\end{equation}
where $B_{j,H}$ is the channel bandwidth between UAV $j$ and HAP $H$. 
$p^{tr}_j$ is the  maximum transmission power of UAV $j$. 
Then, we obtain the time cost to transmit data $\phi_i$ from UAV $j$ to HAP $H$ as
\begin{equation}
T^{tr}_{i,j,H}=\frac{\phi_iz_{i,j,H}}{R_{j,H}}.
\end{equation}

\subsection{Computation Model}
Let $\lambda_j$ denote the computing resource consumed on UAV $j$ to process 1bit data, 
and $C_j$ is the computation capability of UAV $j$ \cite{10}. Then, the delay of TD $i$ 
computed by UAV $j$ is
\begin{equation}
T^{cp}_{i,j}=\frac{\phi_i y_{i,j}\lambda_j}{C_j}.
\end{equation}
Similarly, the delay of TD $i$ computed by HAP $H$ is
\begin{equation}
T^{cp}_{i,j,H} =\frac{\phi_iz_{i,j,H}\lambda_H}{C_H},
\end{equation}
where $\lambda_H$ indicates the computing resource cost of HAP $H$ to handle 1bit data, and  $C_H$ 
represents the computation capability of HAP $H$.
 Hence, the total delay $T_i$ for handling data $\phi_i$ is
\begin{equation}
T_i=\sum\limits_{j=1}^{J}\left(T^{tr}_{i,j}+T^{tr}_{i,j,H}+T^{cp}_{i,j}+T^{cp}_{i,j,H}\right).
\end{equation}

\subsection{Energy Model}
\subsubsection{UAV-based Energy Model}
The energy consumption $E_j$ of UAV $j$ mainly consists of three parts: the basic cost $E^{bas}_j$,
communication cost $E^{tr}_j$, and  computing cost $E^{cp}_j$, i.e.,
\begin{align}
\notag E_j&=E^{bas}_j+E^{tr}_j+E^{cp}_j\\
&=E^{bas}_j+P^{tr}_j \sum\limits_{i=1}^{I}T^{tr}_{i,j,H}+\beta_U C_j^3 \sum\limits_{i=1}^{I}T^{cp}_{i,j},
\end{align}
where $P^{tr}_j$ denotes the transmission power from UAV $j$ to HAP $H$, and  $\beta_U$ indicates
the energy consumption coefficient depending on the chip structure of MEC processors on UAVs.

\subsubsection{HAP-based Energy Model}
The energy consumption $E_H$ of HAP $H$ is composed of the basic cost $E^{bas}_H$
and computation cost $E^{tr}_H $. Let $\beta_H$ denote the energy consumption coefficient depending 
on the chip structure of MEC processors on HAP $H$. Accordingly, $E_H$ can be calculated as
\begin{align}
E_H=E^{bas}_H+E^{cp}_H
=E^{bas}_H+\beta_H C_H^3 \sum\limits_{i=1}^{I}\sum\limits_{j=1}^{J}T^{cp}_{i,j,H}.
\end{align}
\vspace{-0.5cm}

\subsection{Uncertainty Set Construction}
In most cases, the data sizes of tasks are uncertain and their probability distributions are unknown. To 
 enhance the robustness of the model, we denote the 
probability distribution of $\phi_i$ as $\mathbb{P}_i$ and  
the reference distribution of $\phi_i$ is $\mathbb{P}^0_i$, derived from the historical
data. Then, based on $\mathbb{P}^0_i$, 
we define the uncertainty set $\mathscr{D}_i$ as
\begin{equation}
\mathscr{D}_i=\{\mathbb{P}_i|d(\mathbb{P}^0_i,\mathbb{P}_i)\leq \epsilon \}\label{Di},
\end{equation}
where $d(\mathbb{P}^0_i,\mathbb{P}_i)$ is a predefined distance measure between $\mathbb{P}^0_i$  
and $\mathbb{P}_i$,
and $\epsilon$ is the corresponding tolerance value. 
The sample space $\Omega$ contains $K$ possible discrete values of the task
volume, i.e., $\Omega = \{\phi^k| \forall k=1,2,...,K\}\label{uncertainty set}$. Besides, each $\phi_i$ is supposed to have the same sample space $\Omega$ and  follows its respective probability distribution $\mathbb{P}_i$. For a series of obtained historical data, whose number is denoted as $Q$, 
the reference distribution is expressed as 
$\mathbb{P}^0_i=\{p_{i,1}^0,p_{i,2}^0,...,p_{i,K}^0\}$.
Let $p_{i,k}^0 \in \mathbb{P}^0_i$ indicate the probability of $\phi^k$ in the reference distribution, i.e.,
\begin{equation}
p_{i,k}^0=\frac{\sum^Q_{q=1}\delta^k(\phi_i)}{Q},\forall k=1,2,...,K.
\end{equation}
Wherein, if $d^k\leq\phi_i<d^{k+1}$, $\delta^k(\phi_i)$$=$$1$, and  otherwise $\delta^k(\phi_i)$$=$$0$. 
The number of historical data whose data size within interval 
$[d^k, d^{k+1})$ is denoted as $\sum^Q_{q=1}\delta^k(\phi_i)$.

Moreover, we use $L_1$ norm metric to quantify the distance between $\mathbb{P}^0_i$ and $\mathbb{P}_i$,
which is defined as
\begin{equation}
d_{L_1}(\mathbb{P}^0_i,\mathbb{P}_i)=||\mathbb{P}^0_i - \mathbb{P}_i||_1=\sum\limits_{k=1}^K|p_{i,k}-p_{i,k}^0|\label{L1},
\end{equation}
where $p_{i,k}$ denotes the probability of $\phi^k$ in the distribution of $\phi_i$.
 The convergence rate of $d_{L_1}\left(\mathbb{P}^0_i, \mathbb{P}_i\right)$ (denotes as $\nu$) \cite{pan2017data} is
derived from
\begin{equation}
Pr\left(d_{L_1}(\mathbb{P}^0_i,\mathbb{P}_i)\leq \epsilon\right)\geq 1-2Kexp\left(-\frac{2Q\epsilon}{K}\right)=\nu,
\end{equation}
where $Pr\left(A\right)$ indicates the probability of $A$ and  $\epsilon$ denotes the tolerance value:
\vspace{-0.2cm}
\begin{align}
\epsilon=\frac{K}{2Q}\ln\left(\frac{2K}{1-\nu}\right).
\end{align} 

\vspace{-0.88cm}
\subsection{Problem Formulation}
\vspace{-0.2cm}
On the basis of the constructed uncertainty sets, we formulate the computation offloading problem in the AAN 
to minimize the total delay into the following mixed integer minimax optimization problem:
\vspace{-0.1cm}
\begin{align}
\mathbf{P0}\textrm{:}\;&\underset{\boldsymbol{x},\boldsymbol{y},\boldsymbol{z}}{\textrm{min}}\underset{\mathbb{P}_i}{\textrm{max}}\sum\limits_{i=1}^{I}\mathbb{E}_{\mathbb{P}_i}\left(T_i\right)\\
\textrm{s.t.}\;
&\sum\limits_{j=1}^{J}x_{i,j}=1, \forall i\in\mathcal{I}, \label{fuzhu1}\\
&\sum\limits_{i=1}^{I}x_{i,j}\leq N_u, \forall j\in \mathcal{J}, \label{fuzhu2}\\
&\sum\limits_{i=1}^{I}\sum\limits_{j=1}^{J}z_{i,j,H}\leq N_H, \forall j\in \mathcal{J}, i\in \mathcal{I}, \label{fuzhu6}\\
&y_{i,j}+ z_{i,j,H} = x_{i,j},\forall i\in\mathcal{I}, j\in \mathcal{J},\label{fuzhu5}\\
&E_j\leq E_j^{max}, \forall j\in\mathcal{J}, \label{limited energy of UAV}\\
&E_H\leq E_H^{max},\forall i\in\mathcal{I}, j\in \mathcal{J}, \label{limited energy of HAP}\\
&\mathbb{P}_i\in{\mathscr{D}_i}, \forall i\in\mathcal{I},\label{p in d}\\
&x_{i,j}\in\{0,1\},\forall i\in\mathcal{I}, j\in \mathcal{J},\label{01 x}\\
&y_{i,j}\in\{0,1\},\forall i\in\mathcal{I}, j\in \mathcal{J},\label{01 y}\\
& z_{i,j,H}\in\{0,1\},\forall i\in\mathcal{I}, j\in \mathcal{J},\label{01 z}
\end{align}
where $\mathbb{E}_{\mathbb{P}_i}\left(T_i\right)$ represents the expected value of $T_i$ under probability distribution $\mathbb{P}_i$, 
$\boldsymbol{x}= \{x_{i,j},\forall i\in\mathcal{I}, j\in \mathcal{J}\}$, $\boldsymbol{y} = \{y_{i,j},\forall i\in\mathcal{I}, j\in \mathcal{J}\}$, and  $\boldsymbol{z}= \{ z_{i,j,H},\forall i\in\mathcal{I}, j\in \mathcal{J}\}$. 
Constraint (\ref{fuzhu1}) ensures that a task can be offloaded by only one UAV. Constraints (\ref{fuzhu2}) and (\ref{fuzhu6}) 
ensure that the accommodated devices by a UAV or HAP $H$ can not exceed
the quota $N_u$ of UAV $j$ or the quota $N_H$ of HAP $H$. 
Constraint (\ref{fuzhu5}) indicates the data flow conservation at a UAV. 
Constraints (\ref{limited energy of UAV}) and (\ref{limited energy of HAP}) denote the energy budget of UAV $j$ and HAP $H$. 
Constraint (\ref{p in d}) implies that probability distribution $\mathbb{P}_i$ belongs to uncertainty set ${\mathscr{D}_i}$.
Considering that the objective function is an NP-hard mixed integer minimax optimization, with the uncertainty sets in constraints, 
problem $\mathbf{P0}$ is challenging to handle in practice.

\vspace{-0.5cm}
\section{ALGORITHM DESIGN}\label{section3}
\vspace{-0.1cm}
In this section, we reformulate problem $\mathbf{P0}$ and design an algorithm to obtain the final solution.
Firstly, by discretizing the sample space into $K$ levels as mentioned in Section \ref{uncertainty set},
 we reformulate problem $\mathbf{P0}$ as 
\begin{align}
\mathbf{P1}\textrm{:}\;&\underset{\boldsymbol{x},\boldsymbol{y},\boldsymbol{z}}{\textrm{min}}\underset{\mathbb{P}_i}{\textrm{max}}\sum\limits_{i=1}^{I}\sum\limits_{k=1}^K p_{i,k} T_i^k\\
\notag \textrm{s.t.}\;
&(\ref{fuzhu1})-(\ref{fuzhu5}), (\ref{01 x})-(\ref{01 z})\\
&E_j^{bas}+\sum\limits_{k=1}^Kp_{i,k}\phi^k\left(E_j^{tr}+E_j^{cp}\right)\leq E_j^{max}, \forall j\in \mathcal{J} \label{fuzhu3},\\
&E_H^{bas}+\sum\limits_{k=1}^Kp_{i,k}\phi^kE_H^{cp}\leq E_H^{max},\forall i\in\mathcal{I}, j\in \mathcal{J}, \label{fuzhu4}\\
&\sum\limits_{k=1}^{K}p_{i,k}=1, \forall i\in\mathcal{I}, \label{con:pik=1}\\
&\sum\limits_{k=1}^K|p_{i,k}-p_{i,k}^0|\leq \epsilon, \forall i\in\mathcal{I}, \label{con:pik<}
\end{align}
where
\begin{align}
\notag &T_i^k=\sum\limits_{j=1}^J\left(T^{tr,k}_{i,j}+T^{tr,k}_{i,j,H}+T^{cp,k}_{i,j}+T^{cp,k}_{i,j,H}\right)\\
&=\sum\limits_{j=1}^J\left(\frac{\phi^kx_{i,j}}{R_{i,j}}+\frac{\phi^kz_{i,j,H}}{R_{j,H}}+\frac{\phi^k y_{i,j}\lambda_j}{C_j}+\frac{\phi^kz_{i,j,H}\lambda_H}{C_H}\right).
\end{align}
Constraint (\ref{con:pik=1}) is the basic constraint of probability, and  
constraint (\ref{con:pik<}) indicates that probability distribution $\mathbb{P}_i$ belongs to  uncertainty set $\mathscr{D}_i$
 according to (\ref{Di}) and (\ref{L1}).

To address problem $\mathbf{P1}$, we temporarily assume $\mathbb{P}_i$ is fixed, and  relax
the integer variables $\boldsymbol{x}$, $\boldsymbol{y}$ and $\boldsymbol{z}$ into 
continuous variables $\boldsymbol{\tilde{x}}$, $\boldsymbol{\tilde{y}}$ and $\boldsymbol{\tilde{z}}$. 
Then, the outer minimization problem is
transformed as
\begin{align}
\mathbf{P2}\textrm{:}\;&\underset{\boldsymbol{\tilde{x}},\boldsymbol{\tilde{y}},\boldsymbol{\tilde{z}}}{\textrm{min}}\sum\limits_{i=1}^{I}\sum\limits_{k=1}^K p_{i,k} T_i^k\\
&\textrm{s.t.}\;
\notag (\ref{fuzhu1})-(\ref{fuzhu5}), (\ref{fuzhu3}), (\ref{fuzhu4})\\
&\tilde{x}_{i,j} \in [0,1], \forall i\in\mathcal{I}, j\in \mathcal{J},\\
&\tilde{y}_{i,j} \in [0,1], \forall i\in\mathcal{I}, j\in \mathcal{J},\\
&\tilde{z}_{i,j,H} \in [0,1], \forall i\in\mathcal{I}, j\in \mathcal{J}.
\end{align}

To solve the variables $\mathbb{P}_i\in{\mathscr{D}_i}$ in the inner layer, the
minimization problem is transformed into its dual problem $\mathbf{P3}$:
\begin{align}
\mathbf{P3}\textrm{:}\;&\underset{\boldsymbol{\alpha}}{\textrm{max}}
\notag \sum\limits_{i=1}^{I} \alpha^{(\ref{fuzhu1})}_i-\sum\limits_{j=1}^J\bigl(N_u \alpha^{(\ref{fuzhu2})}_j+E^{max}_j \alpha^{(\ref{fuzhu3})}_j-E^{bas}_j \alpha^{(\ref{fuzhu3})}_j\bigr)\\
&-\sum\limits_{i=1}^{I}\sum\limits_{j=1}^{J}\bigl(E^{max}_H \alpha^{(\ref{fuzhu4})}_{i,j}-E^{bas}_H \alpha^{(\ref{fuzhu4})}_{i,j}+N_H \alpha^{(\ref{fuzhu6})}_{i,j}\bigr)\\
\textrm{s.t.}\;
&\alpha^{(\ref{fuzhu1})}_i-\alpha^{(\ref{fuzhu2})}_j-\alpha^{(\ref{fuzhu5})}_{i,j}\leq \sum\limits_{k=1}^{K}p_{i,k}\phi^k R^{-1}_{i,j}, \forall i\in\mathcal{I},  j\in \mathcal{J}, \label{con of x}\\ \notag
&\left(\sum\limits_{k=1}^{K}p_{i,k}\phi^k\right)\left(-\beta_U C^2_j\lambda_j\alpha^{(\ref{fuzhu3})}_j\right)+\alpha^{(\ref{fuzhu5})}_{i,j}\\ 
&\leq\sum\limits_{k=1}^{K}p_{i,k}\phi^k \lambda_j C_j^{-1}, \forall i\in\mathcal{I},  j\in \mathcal{J}, \label{con of y}\\ \notag
&\left(\sum\limits_{k=1}^{K}p_{i,k}\phi^k\right)\left(-p^{tr}_j R^{-1}_{j,H}\alpha^{(\ref{fuzhu3})}_j- \beta_H C^2_H\lambda_H\alpha^{(\ref{fuzhu4})}_{i,j}\right)+\alpha^{(\ref{fuzhu5})}_{i,j}\\
&-\alpha^{(\ref{fuzhu6})}_{i,j}
\leq \sum\limits_{k=1}^{K}p_{i,k}\phi^k \left(\lambda_H C_H^{-1}+R^{-1}_{j,H}\right), \forall i\in\mathcal{I},  j\in \mathcal{J}, \label{con of z}\\ 
&\alpha^{(\ref{fuzhu2})}_j \geq 0,\forall j\in \mathcal{J}, \label{con of fuzhu2}\\
&\alpha^{(\ref{fuzhu6})}_{i,j}\geq 0, \forall i\in\mathcal{I} ,j\in \mathcal{J}, \label{con of fuzhu6}\\
&\alpha^{(\ref{fuzhu5})}_{i,j}\geq 0, \forall i\in\mathcal{I} ,j\in \mathcal{J}, \label{con of fuzhu5}\\
&\alpha^{(\ref{fuzhu3})}_j \geq 0, \forall j\in \mathcal{J}, \label{con of fuzhu3}\\
&\alpha^{(\ref{fuzhu4})}_{i,j} \geq 0,\forall i\in\mathcal{I} ,j\in \mathcal{J}, \label{con of fuzhu4}
\end{align}
where $\boldsymbol{\alpha}= \{\alpha^{(\ref{fuzhu1})}_i, \alpha^{(\ref{fuzhu2})}_j, \alpha^{(\ref{fuzhu6})}_{i,j}, \alpha^{(\ref{fuzhu5})}_{i,j},  \alpha^{(\ref{fuzhu3})}_j, \alpha^{(\ref{fuzhu4})}_{i,j}\}$ are dual variables corresponding to the constraints (\ref{fuzhu1}),
(\ref{fuzhu2}), (\ref{fuzhu6}), (\ref{fuzhu5}), (\ref{fuzhu3})
, and (\ref{fuzhu4}).
Then, problem $\mathbf{P3}$ and the inner maximization operation is combined as
\begin{align}
\mathbf{P4}\textrm{:}\;&\underset{\boldsymbol{\alpha},\boldsymbol{p}}{\textrm{max}}
\notag \sum\limits_{i=1}^{I}\sum\limits_{j=1}^J \alpha^{(\ref{fuzhu1})}_i-N_u \alpha^{(\ref{fuzhu2})}_j-E^{max}_j \alpha^{(\ref{fuzhu3})}_j\\&+E^{bas}_j \alpha^{(\ref{fuzhu3})}_j-E^{max}_H \alpha^{(\ref{fuzhu4})}_{i,j}+E^{bas}_H \alpha^{(\ref{fuzhu4})}_{i,j}-N_H \alpha^{(\ref{fuzhu6})}_{i,j}\\
\notag \textrm{s.t.}\;
&(\ref{con of x})-(\ref{con of fuzhu4}),\\
&\sum\limits_{k=1}^{K}p_{i,k}=1, \forall i\in\mathcal{I},\\
&\sum\limits_{k=1}^K|p_{i,k}-p_{i,k}^0|\leq \epsilon, \forall i\in\mathcal{I},
\end{align}
where $\boldsymbol{p}= \{p_{i,k},\forall i,k\}$.
Problem $\mathbf{P4}$ can be solved 
to obtain $p_{i,k}$ using the optimizer such as Gurobi. Then, problem $\mathbf{P2}$ can be 
handled by the optimizer to obtain continuous $\boldsymbol{\tilde{x}}$, $\boldsymbol{\tilde{y}}$
and $\boldsymbol{\tilde{z}}$. After that, the branch and bound algorithm is adopted
to revert $\boldsymbol{\tilde{x}}$, $\boldsymbol{\tilde{y}}$
and $\boldsymbol{\tilde{z}}$ to integer variables. 
Since the access strategy $\boldsymbol{x}$ should be decided firstly, 
$\boldsymbol{\tilde{x}}$ is reverted to 0 or 1 before $\boldsymbol{\tilde{y}}$
and $\boldsymbol{\tilde{z}}$.
Then, $\boldsymbol{\tilde{y}}$
and $\boldsymbol{\tilde{z}}$ are recovered from the relaxation values. 
The rules for selecting the branch variables are
\begin{align}
x^*=\underset{\tilde{x}_{i,j}\in \boldsymbol{\tilde{x}}}{\textrm{arg max}}\{min\{\tilde{x}_{i,j},1-\tilde{x}_{i,j}\}\}\label{x*}
\end{align}
and
\begin{align}
y^*=\underset{\tilde{y}_{i,j}\in \boldsymbol{\tilde{y}}}{\textrm{arg max}}\{min\{\tilde{y}_{i,j},1-\tilde{y}_{i,j}\}\}\label{y*}.
\end{align}
In short, MDRLOA (i.e., Algorithm \ref{algorithm 1}) is designed for these procedures. 
As step 1 shown, $p_{i,k}$
 is obtained by solving problem $\mathbf{P4}$. Then, we address problem $\mathbf{P2}$
to obtain the continuous values $\boldsymbol{\tilde{x}}$, $\boldsymbol{\tilde{y}}$ and $\boldsymbol{\tilde{z}}$. 
In addition, from steps 4 to 12,
 $\boldsymbol{\tilde{x}}$ is reverted to integer. 
Finally, based on the value of $\boldsymbol{x}$, we can obtain  $\boldsymbol{y}$ and $\boldsymbol{z}$ from steps 14 to 21.
 According to \cite{37},
the computational complexity of reaching $\theta$-optimal
solutions of problem $\mathbf{P2}$ by Gurobi is denoted as 
$\mathcal{O}\left(\sqrt{A_1}ln\left(1/\theta\right)\left(n_1A_1+n_1^2A_1+n_1^3\right)\right)$, 
where $\theta$ is the precision parameter for algorithm in Gurobi to converge to the optimal solution, 
$A_1$ denotes the number of constraints, i.e., $A_1$$=$$6IJ+2J+I$,
and the number of decision variables is $n_1$$=$$3IJ$. 
Similarly, the computational complexity of solving problem $\mathbf{P4}$ is
 $\mathcal{O}\left(\sqrt{A_2}ln\left(1/\theta\right)\left(n_2A_2+n_2^2A_2+n_2^3\right)\right)$, 
where $A_2$$=$$6IJ+2J+2I$ and $n_2$$=$$3IJ+IK+I+2J$.
 Thus, in the worst case, the the computational complexity of MDRLOA is 
\vspace{-1mm}
\begin{equation*}
\begin{aligned}
&\mathcal{O}\Biggl(\sqrt{A_{1}}\ln\left(\frac{1}{\theta}\right)\left(n_{1}A_{1}+n_{1}^{2}A_{1}+n_{1}^{3}\right)+\sqrt{A_{2}}\ln\left(\frac{1}{\theta}\right)\Bigl(n_{2}A_{2} \\
&+ n_{2}^{2}A_{2}+n_{2}^{3}\Bigr)+ 2\sum_{i=1}^{IJ-1}\Bigl(\sqrt{A_{1}+i}\ln\left(\frac{1}{\theta}\right)\Bigl(n_{1}\left(A_{1}+i\right) \\
&+ n_{1}^{2}\left(A_{1}+i\right)+n_{1}^{3}\Bigr)\Bigr)\Biggr)
\end{aligned}
\end{equation*}

\begin{algorithm}[t]
\caption{MDRLOA}

\begin{algorithmic}[1]\label{algorithm 1}

\REQUIRE  Sample space $\Omega$, reference distribution $\mathbb{P}^0_i$ based on a
series of historical data.

\ENSURE $\boldsymbol{x}$, $\boldsymbol{y}$, and  $\boldsymbol{z}$.

\STATE Solve $\mathbf{P4}$ by the
optimizer to obtain $\mathbb{P}_i$.
\STATE Solve $\mathbf{P2}$ by the
optimizer to get relaxed $\boldsymbol{x}$, $\boldsymbol{y}$, and  $\boldsymbol{z}$.

\REPEAT
\STATE Select the branch variable $x^*$ according to (\ref{x*}).
\STATE Add $x^* = 0$, and  all integer $x_{i,j}$ as constraints to $\mathbf{P2}$, and  solve it
to obtain the system latency $L0$.
\STATE Add $x^* = 1$, and  all integer $x_{i,j}$ as constraints to $\mathbf{P2}$, and  solve it
to obtain the system latency $L1$.
\IF{$L0$ < $L1$}

\STATE Update $\mathbf{P2}$ by adding constraint $x^* = 0$.
\ELSE

\STATE  Update $\mathbf{P2}$ by adding constraint $x^* = 1$.

\ENDIF 
\UNTIL all elements in $\boldsymbol{x}$ are integers.
\REPEAT
\STATE Select the branch variable $y^*$ according to (\ref{y*}).
\STATE Add $y^* = 0$, and  all integer $y_{i,j}$ as constraints to $\mathbf{P2}$, and  solve it
to obtain  $L0$.
\STATE Add $y^* = 1$, and  all integer $y_{i,j}$ as  constraints $\mathbf{P2}$, and  solve it
to obtain  $L1$.
\IF{$L0$ < $L1$}

\STATE Update $\mathbf{P2}$ by adding constraint $y^* = 0$.
\ELSE

\STATE  Update $\mathbf{P2}$ by adding constraint $y^* = 1$.

\ENDIF 

\UNTIL all elements in $\boldsymbol{y}$, $\boldsymbol{z}$ are integers.

\end{algorithmic}
\end{algorithm}

\vspace{-1mm}
\section{SIMULATION RESULTS}\label{section4}
\begin{figure}[t]
	\centering
	\subfloat[Real system latency \textit{v.s.} the average size of real data.]
      {
	\includegraphics[width=0.45\linewidth]{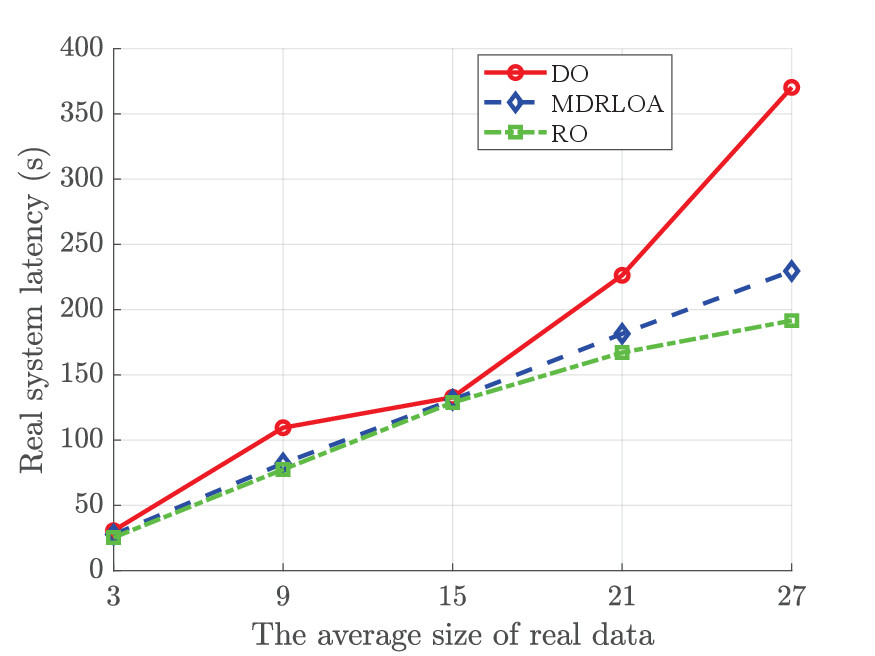}
     \label{fig:1a}
      } 
   \quad 
    \subfloat[Real total energy cost \textit{v.s.} the average size of real data.]
     {
	\includegraphics[width=0.45\linewidth]{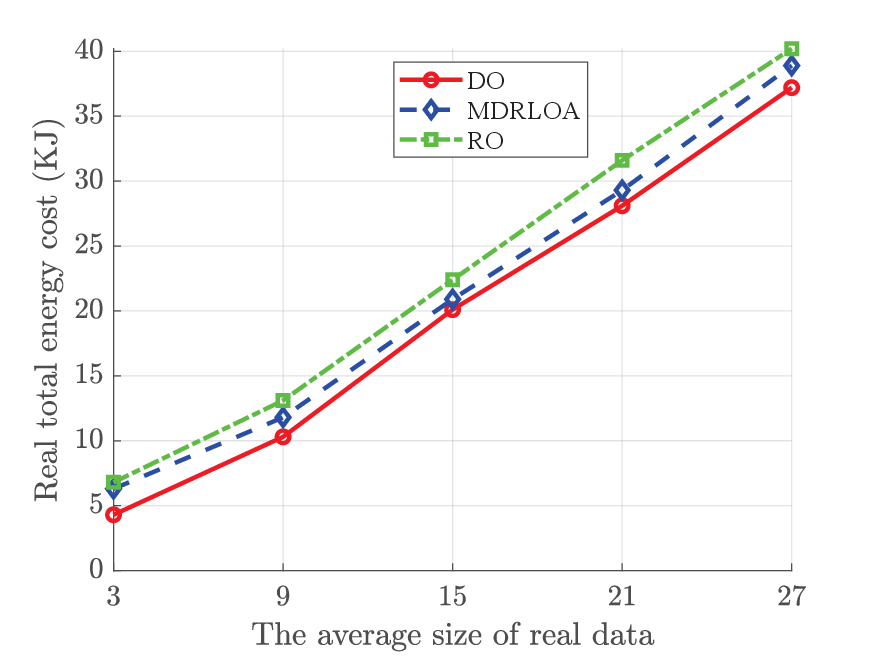}
     \label{fig:1b}
     }
	\caption{Performance of different optimization mechanisms.}
	\label{fig:dif met}
\end{figure}
\begin{figure}[t]
	\centering
	\subfloat[Real system latency \textit{v.s.} number of historical data.]
      {
	\includegraphics[width=0.57\linewidth]{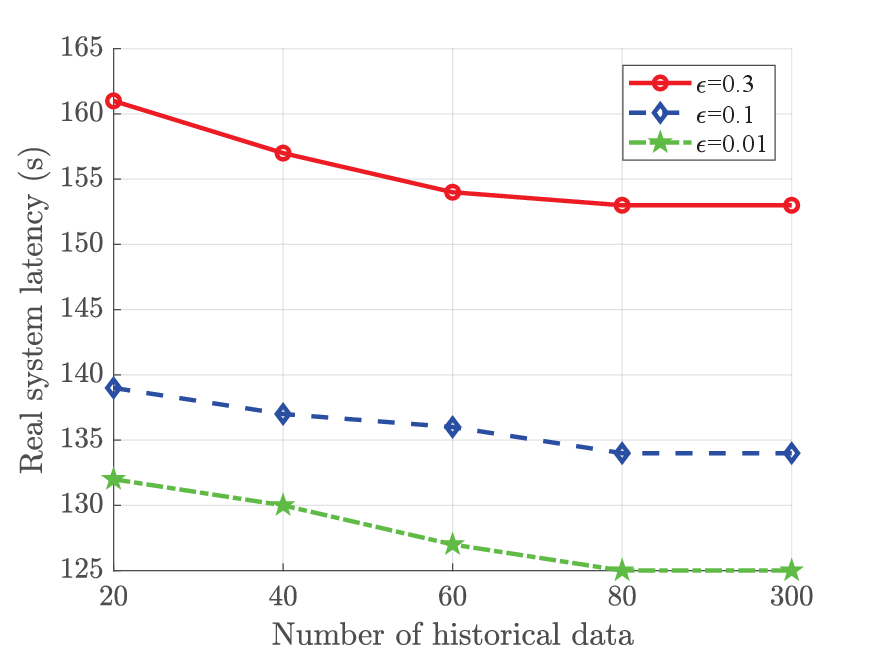}
     \label{fig:2a}
      } 
   \quad 
	\subfloat[Probability distribution function \textit{v.s.} data size.]
      {
	\includegraphics[width=0.57\linewidth]{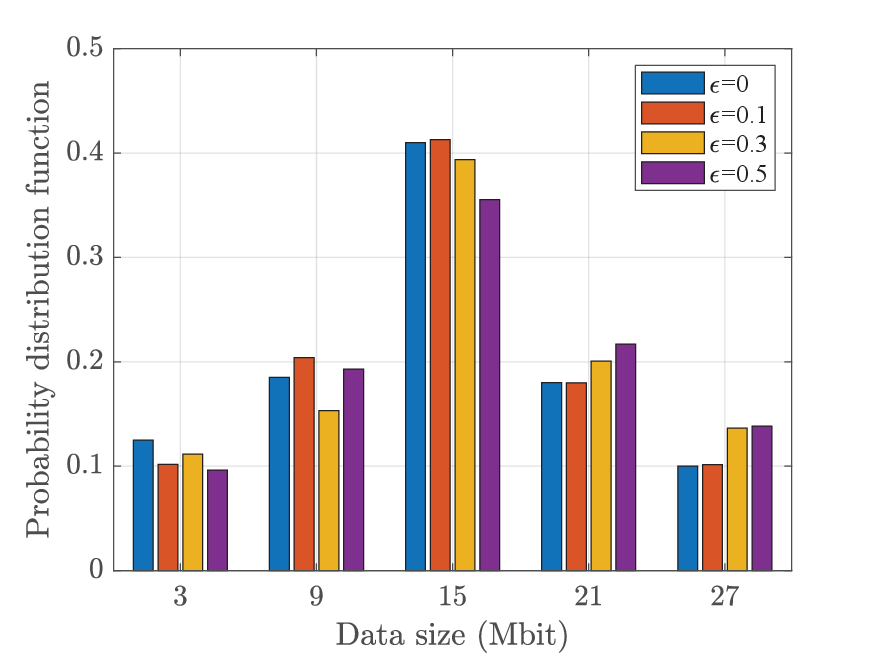}
     \label{fig:2b}
      } 
   \quad 
    \subfloat[Real total energy cost \textit{v.s.} quota of the HAP.]
     {
	\includegraphics[width=0.57\linewidth]{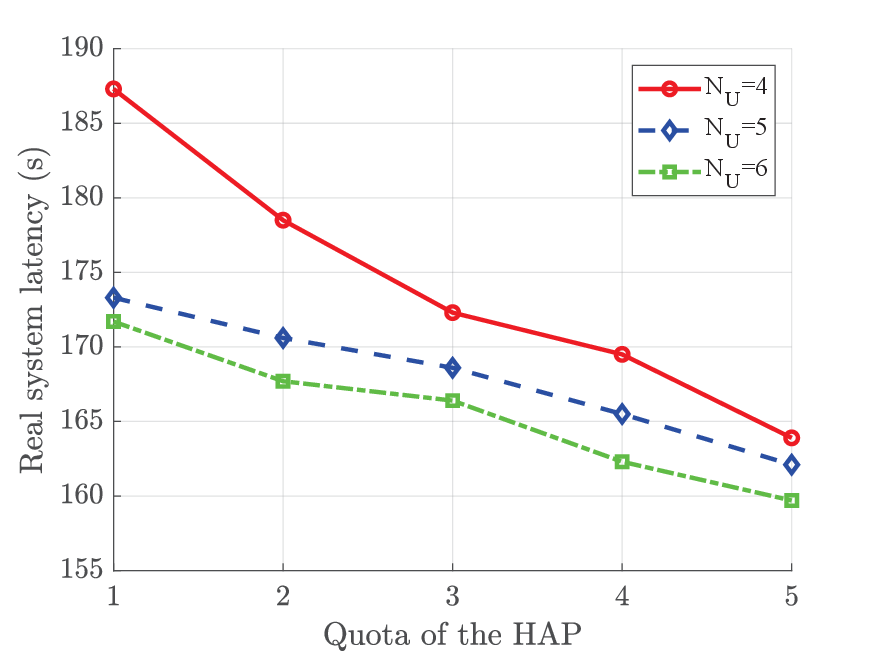}
     \label{fig:2c}
     }
	\caption{Performance of MDRLOA with different parameters.}
	\label{fig:dif par}
\end{figure}

The simulation scenario is set as follows. We assume there is one HAP suspending at the height of $20km$,
 3 UAVs with altitude of $2km$ are randomly
distributed in the area with size of $10km\times10km$, and  several TDs are randomly distributed 
in this area. Note that UAVs are in the coverage of the HAP, and 
TDs are in the coverage of UAVs. The uncertainty set related parameters are set as $K$$=$$5$, and  $\Omega$$=$$\{3, 9, 15, 21, 27\}Mbits$. Besides, following \cite{10} and \cite{50}, the
computation and communication related parameters are set as
$g^0_{i,j}$$=$$-60dB$, $g^0_{j,H}$$=$$-60dB$, $B_{i,j}$$=$$1MHz$, $B_{j,H}$$=$$20MHz$, $\sigma^2$$=$$-100dB$,
 $C_j$$=$$3*10^9cycles/s$, 
$C_H$$=$$5*10^{10}cycles/s$, $\lambda_j$$=$$270cycles/bit$, $\lambda_H$$=$$1,100cycles/bit$, 
$\beta_U$$=$$10^{-28}$, and  $\beta_H$$=$$10^{-28}$. The power related parameters are set as $p^{tr}_i$$=$$0.5W$, $p^{tr}_j$$=$$10W$, 
$E_j^{max}$$=$$100KJ$, and  $E_H^{max}$$=$$1,000KJ$.

In Fig. \ref{fig:dif met}, 
the parameters are set as $Q$$=$$200$, $\epsilon$$=$$0.3$, $N_U$$=$$4$, and  $N_H$$=$$4$.
After obtaining the offloading decision by using DO, MDRLOA and RO,
 we input the real data size to compare the performance of different methods.
 Moreover, in DO, the UAV estimates data sizes to certain values 
in order to design an offloading strategy. 
Here, the average value of basic events in the sample space 
is used as the determined estimate. 
In contrast, data sizes are estimated to the largest values 
for designing an offloading strategy in RO.
In Fig. 2\subref{fig:1a},
 it is noted that the system latency of DO is the highest, followed by MDRLOA, 
and RO lowest.
In Fig. 2\subref{fig:1b},
 the energy cost of RO is the highest, followed by MDRLOA, 
and DO lowest.
It is accounted that, compared with DO, MDRLOA and RO use more computing resource to deal with uncertain tasks,
 leading to lower latency and higher energy cost. 
Thus, compared to DO, the proposed MDRLOA is more conservative. Further,
 MDRLOA is more energy-saving in practical scenarios compared with RO.

In Fig. \ref{fig:dif par}, we explore the performance of MDRLOA for different parameters. 
Fig. 3\subref{fig:2a} shows the system latency of different tolerance values with the number
of historical data.
It is noted that as the number of historical data increases,
 the system latency gradually decreases. Besides, when the tolerance value is larger,
the system latency increases. It is explained from Fig. 3\subref{fig:2b}, 
which shows the estimated distribution used to 
replace the true distribution for different tolerance values. 
 When the tolerance value is larger, the expected value of $\phi_i$ increases.
In other words, the probability distribution of uncertain data size becomes worse.
Thus, with the large tolerance value, the system reliability, and  conservativeness are higher, which
leads to rise in system latency as shown in Fig. 3\subref{fig:2a}.
Moreover, Fig. 3\subref{fig:2c} reveals the system latency associated with the HAP quota 
as well as UAV quotas.
The parameters in this figure are set as $Q$$=$$200$, and  $\epsilon$$=$$0.3$. 
The system latency decreases with the increment of HAP quota, and  UAV quota. It is accounted
for the fact that the system latency is lower when more computing resource is used to complete tasks.

\section{CONCLUSIONS}\label{section5}
\vspace{2mm}
In this paper, we investigate the AAN composed of TDs, UAVs and a HAP
to optimize the system latency with uncertain tasks. To tackle 
the uncertain tasks, we construct the uncertainty set based on $L_1$ norm metrics 
to describe the probability distribution by historical data. Then, 
 a DRO problem is proposed, to minimize the system delay under the worst-case probability distribution.
We further design algorithm MDRLOA
 for effective solutions. Finally, simulation results show that, the robustness of the
proposed algorithm is higher than DO. The system latency of the proposed
algorithm is about $18.8\%$ lower than DO on average. 
Meanwhile, MDRLOA saves about $6.9\%$ energy than RO on average
since RO is more conservative. In short, MDRLOA performs better in practical circumstances.
\vspace{2mm}
\bibliographystyle{IEEEtran}
\bibliography{REFERENCES.bib}
\end{document}